\documentclass[twocolumn,aps,showpacs,floatfix,prc]{revtex4}
\usepackage[dvips]{epsfig}
\newcommand{\snn}{\sqrt{s}_{NN}}
\begin{document}

\title{Charmonium suppression in a baryon rich quark-gluon plasma}\author{Partha Pratim Bhaduri}
\email[E-mail:]{partha.bhaduri@vecc.gov.in}
\author{A. K. Chaudhuri}
\author{Subhasis Chattopadhyay}
\affiliation{Variable Energy Cyclotron Centre,\\ 1/AF, Bidhan Nagar,
Kolkata 700~064, India}     
\date{\today}
\begin{abstract}
 We have investigated the charmonium survival probability, in a high baryon density parton plasma, expected to be produced in nuclear collisions at FAIR. Charmonia are assumed to undergo complete dissociation by color screening, if the in-medium Debye radius becomes comparable to the spatial size of the corresponding bound state. Results indicate a non-trivial dependence of the suppression pattern on the plasma evolution dynamics. A much larger magnitude of suppression is foreseen induced by cold nuclear matter compared to that due to plasma screening.
\end{abstract}  
\pacs{PACS
numbers: 25.75.-q, 25.75.Dw}
\maketitle
 
In relativistic nuclear collisions, $J/\psi$ suppression has long been predicted as an unambiguous and experimentally viable signature to indicate the possible occurrence of phase transition to quark-gluon plasma~\cite{mats}. At SPS, data on $J/\psi$ suppression were collected for Pb+Pb and In+In collisions respectively by NA50~\cite{Ale05} and NA60~\cite{Arn07}  collaborations at a beam kinetic energy ($E_b$) 158 A GeV. Subsequent measurements of charmonium production by NA60 collaboration, in $p+A$ collisions at the same energy~\cite{scompar}, analyzed within Glauber model framework, suggest that the relative charmonium yield in In+In collisions are in accord within errors, with suppression induced by cold nuclear matter. An anomalous suppression of about 25 - 30 $\%$ still remains visible in the most central Pb+Pb collisions. Till date no measurement exists on charmonium production in heavy-ion collisions below the top SPS energy, primarily due to the extremely low production cross sections. This in turn demands accelerators delivering extremely high intensitiy heavy-ion beams and detectors with high rate handling capability. The upcoming Compressed Baryonic Matter (CBM) experiment at FAIR~\cite{Peter}, in GSI, Germany, for the first time, is aiming at the measurement of charmonium production in low energy nuclear collisions, close to the production threshold. In the energy domain, within the reach of the FAIR accelerators ($E_b$=10-40 A GeV), highest possible baryon densities are expected to be produced at the center of the collision zone~\cite{Arsene}. This might lead to a density driven QCD phase transition of the nuclear matter to a baryon rich quark-gluon plasma (QGP). In the present article, we plan to estimate the amount of dissociation of the charmonium states induced by color screening inside a hot baryonic plasma. For this purpose, we have developed a variant of the static geometrical threshold model~\cite{bl96, bl00}. Debye screening mass as a function of temperature and baryon chemical potential $m_D(T,\mu_B)$, in a dynamically evolving plasma is used to decide the fate of different charmonium states implanted in the medium.

Quarkonium suppression in nuclear collisions at SPS, RHIC and LHC has been studied at length in literature~\cite{Blaizot, Chu, Karsch, Ramona, Patra1, Madhukar, Gunji,HSD}, using geometrical screening models.  Based on semi-classical arguments, they assume some sharply defined formation time and a suppression which is total or absent depending on the time spent by the $c\bar{c}$ pair inside plasma being shorter or longer than the Lorentz dilated formation time in the plasma frame (for an alternative quantum mechanical prescription for the evolution of a $c\bar{c}$ wave packet in the plasma see~\cite{Cugnon}). The basic underlying theme, thus is the existence of a characteristic threshold dissociation temperature ($T_d$) or equivalent energy density ($\epsilon_d \simeq T_d^4$), which encloses the plasma volume where the Debye screening length is shorter than the characteristic Bohr radius of a particular bound state. Resonance formation is thus forbidden for all $c\bar{c}$ pairs inside the region at the corresponding resonance formation time $t_R$, in the plasma frame. Competition between $t_R$ and the finite volume and life time of the plasma, leads to the characteristic $p_T$ dependent survival probability at central rapidity region. Finite lifetime sets an upper limit on the $p_T$ at which charmonia are suppressed. The common lore in all such calculations is that the deconfined medium formed in the nuclear collisions developed the thermal properties within a time comparable to the formation time of the primordial $c\bar{c}$ pairs in the collision frame. Hence these $c\bar{c}$ pairs would travel inside the plasma and those which satisfy the screening conditions, would not be able to evolve as the physical bound states. However the situation might be different at FAIR owing to different kinematic conditions. Due to large reduction in collision energies, much longer time will be required for the produced medium to attain thermalization compared to that at higher energies. Though thermalization in nuclear collisions is not yet understood clearly, it is commonly granted that the lower bound on the thermaliztion time can be estimated from the passing time, $t_{d}=2R_{A}/\gamma\beta$, of the two colliding nuclei, where $R_A$ is the nuclear radius and $\gamma$ is the Lorentz contraction factor. At SPS energy ($\snn \simeq 17.3$ GeV), $t_d$ is about 1 fm/c, whereas at FAIR energies $t_d \simeq 3 - 4 fm/c$. On the other hand due to small available energy in the center of mass frame of the collision, the $c\bar{c}$ pairs will be produced with very small $p_T$, maximum of $1 - 2$ GeV/c. Consequently the Lorentz factor $\gamma_{R,i}$  by which the corresponding intrinsic formation time of the $i^{th}$ resonance is dilated in the collision frame, would also be small ($\gamma_{R,i} \simeq 1$, so that $t_{R,i} = \gamma_{R,i}\tau_{R,i} \simeq \tau_{R,i}$). For the charmonium states like $J/\psi$, $\psi'$ or $\chi_c$, the intrinsic formation time is $\tau_{R,i}\approx 1-2$ fm~\cite{time}. Thus it might be reasonable to consider that at FAIR, different charmonium states will be formed in the pre-equilibrium stage and the plasma would encounter the already formed physical bound states rather than their precursors as believed to be the case at higher energies. Once plasma is produced, Debye screening would set in and if the local temperature $T(x)$ and chemical potential $\mu(x)$ inside the plasma are such that the screening radius $r_D(T(x),\mu(x)) \le r_i$, the $i^{th}$ state will melt, where $r_i$ is the rms separation radius of the $i^{th}$ charmonium state, calculated in non-relativistic quarkonium spectroscopy~\cite{satz}. Hence, for an impact parameter {\bf b}, the survival probability at a transverse position {\bf s}, at time $\tau$ can be expressed as
 
\begin{equation}
S_i^{QGP}({\bf b, s}, \tau) = \Theta(r_i - r_D({\bf b, s},\tau))
\label{surv}
\end{equation}

\begin{figure}
  \includegraphics[height=6.5cm,width=6.5cm]{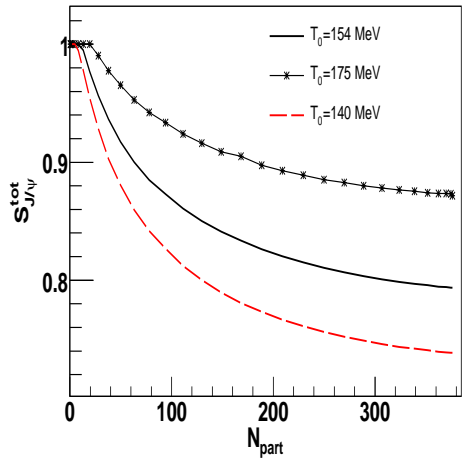}
  \caption{Centrality dependence of the total $J/\psi$ survival probability in a baryon rich QGP medium created in 30 A GeV Au+Au collisions. Dissociation is exclusively due to Debye color screening. Screening masses are evaluated using simple perturbative ansatz. Different curves correspond to different values of $T_0$, an input parameter of the plasma EOS, implemented in our present calculations for characterization of the deconfined phase.}
  \label{F1}
\end{figure}  

Modeling the dissociation by a $\Theta$ function might not be very realistic as it completely ignores the time scale involved in the melting process. In~\cite{bl00}, the authors showed that their theoretical curve obtained using $\Theta$ function could not provide a perfect fit to the NA50 data points with smaller error bars. Fit quality can be improved by considering the gradual suppression mechanism which resulted in smearing of the $\Theta$ function by small amount. Smearing width (an additional model parameter) was fixed by the data. In the same spirit, we smear the $\Theta$ function in Eq.~\ref{surv}; for each resonance state the corresponding smearing width is fixed to $10 \%$ of its corresponding separation radius.  Determination of survival probability thus boils down to the estimation of in-medium Debye screening mass ($m_D(T,\mu_B) = 1/r_D(T,\mu_B)$). At vanishing net baryon density, the color screening theory has been analyzed in quite some detail. Compared to this our knowledge at finite density is rather limited. In leading order perturbation theory, the $\mu_B$ dependence of the screening mass reads as~\cite{mDLO}, $m_D(T,\mu_q) = g(T,\mu_q)T\sqrt{\frac{N_c}{3}+\frac{N_f}{6}+\frac{N_f}{2\pi^2}(\frac{\mu_q}{T})^2}$, where $\mu_q=\mu_B/3$ is the quark chemical potential and the other symbols have their usual meaning. The NLO level estimates of Debye mass suffer from large uncertainties, particularly due to yet unresolved nature of the magnetic screening mass. In the temperature and density ranges expected at FAIR, application of perturbation theory is not beyond doubt. However, in lattice QCD, the density dependence of screening masses has so far been analyzed for 2-flavor QCD and unrealistically large quark masses~\cite{Okacz}. In the present study, we thus desist from using lattice results and restrict ourselves to the LO pQCD estimations only. Running coupling $g(T, \mu_q)$ is obtained from~\cite{Stocker}. Our theory now requires the local $T$ and $\mu_B$ of the fluid, as a function of collision energy and centrality. Recently a number of existing dynamical models based on transport or hydrodynamical equations have been employed to simulate central collisions of gold nuclei in the FAIR energy regime~\cite{Arsene}. For each case, central baryon density ($\rho_B$) and the total energy density ($\epsilon$) are extracted as a function of time. A large degree of mutual agreement among the results ($\rho_B(t)$ and $\epsilon(t)$) of the various models has been observed despite of their very different mean fields and degrees of freedom. This is ascribed to the fact that being mechanical variables $\rho_B$ and $\epsilon$ are subject to the local conservation laws. Since the various dynamical models abide by these basic conservation equations they will tend to give similar results for the corresponding quantities. For our present calculations we resort to the UrQMD model to get $\rho_B(t)$ and $\epsilon(t)$, extracted for a central cell of unit thickness ($\Delta z = 1$ fm) in the longitudinal direction. To account for the spatial non-uniformity of the medium, we set the initial profiles of $\rho_B$ and $\epsilon$ in the transverse plane of the collision, in proportion to the participant density ($n_{part}({\bf {b,s}})$), calculable using Glauber model. Space time dependent density values ($\rho_B$, $\epsilon$) so obtained, can then be simultaneously plugged in some suitable partonic equation of state (EOS) to solve for the corresponding values of local $T$ and $\mu_B$ of the fluid. We here employ phenomenological QGP EOS proposed by Kapusta~\cite{Kapusta:2010ke}. It is so constructed that it matches with the lattice QCD simulation at $\mu_B \approx 0$ and to the known properties of ground state nuclear matter. In the EOS, one input parameter $T_0$ can be identified with the pseudo-critical temperature ($T_c$) at $\mu_B \approx 0$. The critical energy density ($\epsilon_c$), required for deconfinement can then be estimated at $\mu_B \approx 0$. Since $T_c$ as a function of $\mu_B$ is believed to follow a nearly constant density curve, we assume $\epsilon_c$, to remain constant with increase in $\mu_B$. Incorporation of $\epsilon_c$ endows the plasma with finite space-time extent. We now have all the ingredients to calculate the centrality dependence of the $J/\psi$ suppression pattern due to screening in the QGP medium. In practice, it has been found that only about $60\%$ of the observed $J/\psi$ originate directly in hard collisions while $30\%$ of them come from the decay of $\chi_c$ and $10\%$ from the $\psi^{'}$. Hence, the total survival probability of $J/\psi$ becomes, $ S^{tot}$=$0.6S_{J/\psi}+0.3S_{\chi_c}+0.1S_{\psi^{'}}$. Inclusive survival probability for different states, can be obtained by integrating Eq.~\ref{surv} over space-time. To do so we distribute $J/\psi$ in the transverse plane, following the transverse density of binary collisions $n_{coll}({\bf b, s})$, as obtained in Glauber model. Screening effects remain operational on a bound state, over the time it spends inside the plasma. The lower limit of the screening time is assumed to coincide with the thermalization time of the medium ($t_{th}$). Relaxation to local equilibrium cannot occur earlier than a certain time needed for the Lorentz contracted nuclei to pass through each other. Following~\cite{Bravina}, we take $t_{th} =(\frac{2R_A}{\gamma\beta}+\frac{\Delta z} {2\beta})$, which sets the lower bound of thermalization time. A charmonium state of mass $m_{\psi}$, inserted at a point ${\bf r}$, in the transverse plane ($z=0$) of the plasma, with  velocity  ${\bf v}$ will travel a distance d ($d = - rcos \phi + \sqrt{R^2-r^2(1-cos^2 \phi)}$), in the time interval $t_b = M_T d /p_T$, before it escapes from the partonic system of transverse  extension $R$, $\phi$  being the angle between the vectors ${\bf r}$ and ${\bf v}$ and $M_T (M_T=\sqrt{(p_T^2+m^2_{\psi}})$ denotes the transverse mass. Hence if $t_I$ denotes the plasma extinction time, minimum between the two time values, $t_I$ and $t_b$ should set the upper limit of duration a bound state suffers dissociation due to screening. However as pointed out earlier, the quarkonia produced at FAIR energies in the mid-rapidity ($p_z \simeq 0$) are expected to have very small $p_T$.  Experiments measuring $J/\psi$ production cross sections in $p+A$ collisions at BNL-AGS~\cite{AGS} and CERN-PS~\cite{PS} in the same energy domain as FAIR, reported the $p_T$ distrbution to be described $p_T \simeq exp(-6p_T)$ or $p_T \simeq exp(-ap_T^2)$ with $a=1.6$ GeV$^{-2}$ respectively. Such anzatzes correspond to an average transverse momentum of $<p_T> \simeq 0.17$ GeV (exponential distribution) or $<p_T> \simeq 0.45$ GeV (Gaussian distribution). For such small $p_T$, the charmonia are found to remain inside the plasma, as long as the plasma is alive. This would essentially result $p_T$ independent survival probability.

\begin{figure}
\includegraphics[height=6.5 cm,width=6.5 cm]{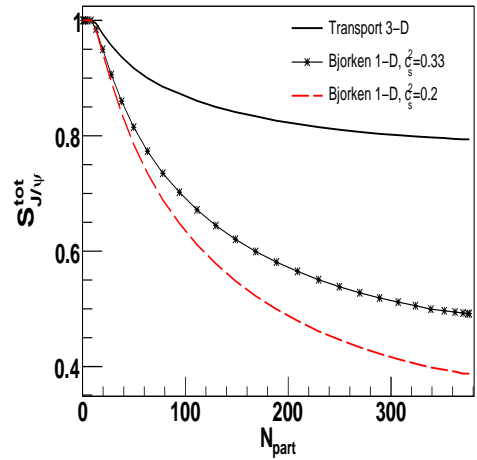}
\caption{Sensitivity of $J/\psi$ survival probability, suffering dissociation from Debye screening, to the plasma expansion dynamics, in 30 A GeV Au+Au collisions. Curves are generated for $N_f=3$ and $T_0=154$ MeV. Fluid is assumed to expand with and without transverse expansion. In case of longitudinal Bjorken type expansion, two different $c_s^2$ values are considered. Suppression effects are more stronger in absence of the transverse motion. Incorporation of transverse dynamics helps the partonic fluid to dilute and cool at a much faster rate and thus diminish the dissociation effects.}
  \label{F2}
\end{figure}

\begin{figure}
  \includegraphics[height=6.5cm,width=6.5 cm]{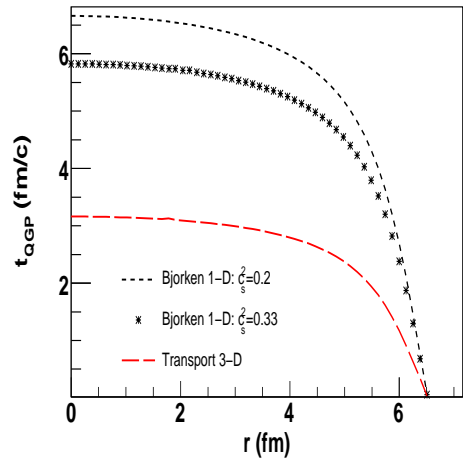}
\caption{Dependence of the plasma life time on the medium expansion dynamics. Curves show the variation of the plasma life time as a function of the transverse co-ordinate $r$, in 30 A GeV central Au+Au collisions. Incorporation of transverse dynamics significantly reduceses the duration of the deconfined phase.}
  \label{F2a}
\end{figure} 

 In Fig.~\ref{F1}, we present the inclusive total $J/\psi$ survival probability as a function of $N_{part}$ in 30 A GeV Au+Au collisions. In order to estimate the sensitivity of the screening dissociation on the plasma EOS, calculations are made with three different values of $T_0$, which also corresponded to variation of $\epsilon_c$. Guided by the current state-of-the-art lattice simulations of $T_c$~\cite{Bazavov}, we have chosen $T_0=154$ MeV. The lowest value ($T_0=140$ MeV) arises from the hope that $T_c$ cannot be less than the pion rest mass. The highest value ($T_0=175$ MeV) is inspired from~\cite{sgupta}. Suppression is indeed found to be sensitive to the choice $T_0$. A higher $T_0$ in turn results in higher $\epsilon_c$ for deconfinement and thus shrinks the spatio-temporal extent of the plasma. All three curves are generated for $N_f=3$ and the flavor dependence of the suppression pattern is rather very small and not shown explicitly.

It would also be interesting to find out the sensitivity of the suppression pattern on the medium expansion dynamics. Most of the existing calculations so far, generally assume the plasma to expand following Bjorken 1-D boost invariant scaling solutions. It is basically driven by the expectation that the suppression occurs before the transverse expansion could set in. In our present calculations, we have followed the time evolution of the densities from UrQMD which includes the full 3-D expansion of the fireball. For comparison, we have also calculated the suppression for a case where the fluid dynamics is governed by Bjorken 1-D hydrodynamical expansion. For such cases the net baryon and the energy densities would evolve as $\rho_b\tau = constant$ and $\epsilon\tau^{1+c_s^2} = constant$, where $c_s$ denotes the speed of sound in the plasma. For ideal massless relativistic gas $c_s^2 = 0.33$. For a massive interacting system $c_s^2$ would be less than that.  In Fig.~\ref{F2}, we have shown comparative suppressions between 3-D and 1-D expansions of the partonic fluid. Magnitude of suppression exhibits a strong dependence on the medium expansion dynamics, with larger suppression occuring in absence of transverse expansion. This can be clearly understood by looking at Fig.~\ref{F2a}, where we have plotted the variation of plasma life time as a function of transverse co-ordinate $r$, in 30 A GeV central Au+Au collisions. In absence of transverse expansion, plasma expands slowly and hence screening would operate for a longer time. Expansion rate in such cases, would be controlled by the value of $c_s^2$. Among the three cases considered, slowest possible expansion occurs for $c_s^2=0.2$, giving largest suppression effects. On the other hand presence of the transverse expansion accelerates the rate of expansion thereby shortens the plasma life, which would increase the chances for $J/\psi$ to escape the dissociation. A static plasma would then result in maximum possible suppression due to screening effects.
\begin{figure}
 \includegraphics[height=6.5cm,width=6.5 cm]{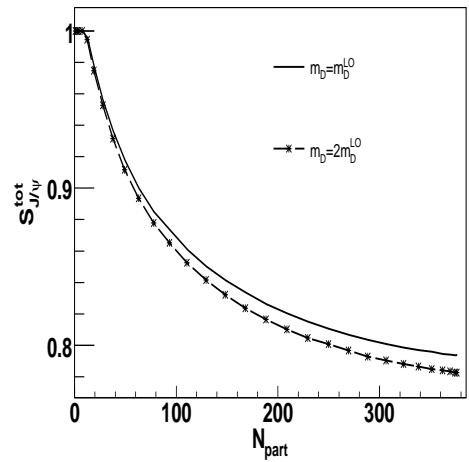}
\caption{Sensitivity of the $J/\psi$ suppression on the non-perturbative effects in the Debye screening mass. Centrality dependence of total inclusive survival probability of $J/\psi$, suffering dissociation due to plasma screening, is calculated for 30 A GeV Au+Au collisions. The upper curve represents the generic case where in-medium screening masses are evaluated using leading order perturbative QCD. The lower curve is generated for a case where the screening mass is twice that of LO pQCD, a factor arbitrarily chosen to account for non perturbative effects. Results show that the non perturbative contribution to the Debye mass induce a very little effect on the observed $J/\psi$ inclusive suppression pattern.}
  \label{F2b}
\end{figure} 
 
While discussing the effects of plasma screening, it might also be relevant to estimate the sensitivity of the suppresion pattern on the evaluation of screening mass. In our model calculations, we have used a simple leading order perturbative ansatz for estimation of Debye screening mass, due to want of suitable calculations at finite density. At $\mu_B \approx 0$, lattice studies~\cite{Okacz} indicate that for $T \le 1.5T_c$, LO pQCD estimations of $m_D$ are underestimated by a factor $A \approx 1.4$. Such non-perturbative effects might also be important at FAIR energy domain. Again if Debye masses, at finite $\mu_B$, are evaluated at next-to-leading order level following the prescriptions in~\cite{Hong}, they come out to be on the average about $1.9$ times larger than the LO estimates. To explicitly check the effect of non-perturbative corrections to the screening masses on the observed $J/\psi$ suppression pattern, we have calculated the total inclusive survival probability for an arbitrarily chosen large non-perturbative correction factor $A=2$, as shown in Fig.~\ref{F2b}. The suppression merely increases at the most by $2\%$ in near central collisions. Such small changes would be too meagre to affect the overall suppression and possibly will not be detected within finite experimental resolution.

\begin{figure}
 \includegraphics[height=6.5 cm,width=6.5 cm]{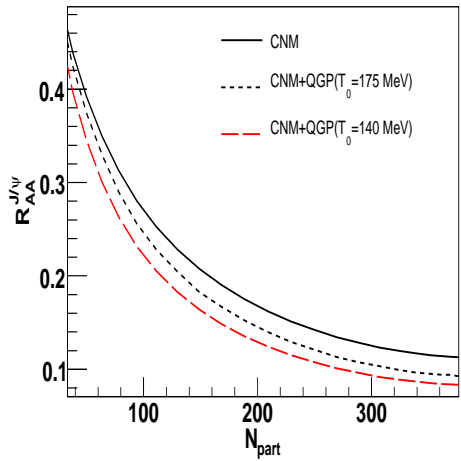}
\caption{ Model predictions for the centrality dependence of the nuclear modification factor for $J/\psi$ ($R_{AA}^{J/\psi}$) in 30 A GeV Au+Au collisions. $J/\psi$' s are mostly suppressed due to CNM effect. Formation of the deconfined matter enhances the dissociation, maximum by a factor of $15 - 25\%$ depending on the choice of $T_0$, an input parameter of the plasma EOS.}
  \label{F3}
\end{figure}  

Up till now we have discussed the $J/\psi$ suppression resulting only from the screening effects inside the plasma. In FAIR energy collisions, charmonium production predominantly occurs through the initial hard nucleon-nucleon collisions. Only those primordial $J/\psi$ which would be able to survive after the so called cold nuclear matter (CNM) effects will subject to the Debye screening effects provided the nuclear collisions eventually lead to the formation of the plasma phase. Hence for a quantitative comparison with the experimental data, we calculate the so called nuclear modification factor ($R_{AA}$) for $J/\psi$ in 30 A GeV Au+Au collisions. Since the CNM effects and the color screening become operative at different time intervals one is allowed to write $R_{AA}^{J/\psi}=R_{AA}^{CNM}\times S_{J/\psi}^{tot}$. The $J/\psi$ production and its suppression in the nuclear medium for 30 A GeV Au+Au collisions, is calculated following the adapted version of the QVZ model~\cite{Qui}, as reported in~\cite{partha2}. The model assumes $J/\psi$ production in hadronic collisions, to be a factorisable two step process:(i) formation of $c\bar{c}$ pair, accounted by perturbative QCD and (ii) formation of $J/\psi$ meson from the $c\bar{c}$ pair, which is non-perturbative in nature and conveniently parameterized. Different parametric forms have been formulated for the transition probability, $F_{c\bar{c} \rightarrow  J/\psi}(q^2)$, of a $c\bar{c}$ pair with relative momentum square $q^2$ to evolve into a physical meson, following the existing models of color neutralization. Out of them the power law form ($F^{\rm (P)}(q^2)$) bearing the essential features of the Color-Octet~\cite{Octet} models has been found earlier to describe the latest $J/\psi$ production cross section data at SPS, in p+A~\cite{partha1}, as well as in In+In and Pb+Pb collisions~\cite{partha2} reasonably well. Hence in the present calculation we opt for the power law form.  The CNM effects incorporated in our model calculations include the nuclear modification of the parton densities (shadowing effects), at the initial stage and the final state dissociation of the pre-resonance $c\bar{c}$ pairs. In our analysis, leading order MSTW2008~\cite{MSTW} set is used for free proton pdf and EPS09~\cite{EPS09} interface for the ratio $R_i(A,x,Q^2)$, that converts the free-proton distributions for each parton $i$, $f_i^{p}(x,Q^2)$, into nuclear ones, $f_i^A(x,Q^2)$. In nuclear collisions, parton densities are modified both inside projectile and target nuclei. Depending on the collision geometry, either the halo or the core of the nuclei gets mainly involved, and the resulting shadowing effects appear to be more important in the core than in the periphery. Assuming shadowing is proportional to the local nuclear density, approximated by two parameter Woods-Saxon density distribution, shadowing factors are evaluated as a function of collision centrality. Within QVZ approach, in contrast to the conventional Glauber model framework, the final state dissociation of the pre-resonant $c\bar{c}$ pairs, is employed through multiple scattering of the heavy-quark pair in the nuclear medium.  It is assumed that once produced, the nascent $c\bar{c}$ pairs interact with nuclear medium and gain relative square momentum at the rate of $\varepsilon^2$ per unit path length inside the nuclear matter. As a result, some of the $c\bar{c}$ pairs can gain enough  momentum to cross the threshold to become open charm mesons, leading  to the reduction in $J/\psi$ yield compared to the nucleon-nucleon collisions. For the power law parameterization of transition probability, the corresponding value of $\varepsilon^2$, extracted from the analysis of p+A collision data~\cite{partha1}, exhibited non-trivial beam energy dependence. Lower be the beam energy, higher is the value of $\varepsilon^2$ implying larger nuclear dissociation of the charmonium precursor states. In the present work we have employed the $\varepsilon^2(E_b)$ value previously parametrized in~\cite{partha1}. Results are shown in Fig.~\ref{F3}, for nuclear modification factor, with and without QGP effect. As evident from the figure, the overwhelming contribution to the observed $J/\psi$ suppression comes from the CNM effects. In the kinematic regime probed at FAIR energy collisions, nuclear densities of both the gluons as well as valence quarks exhibit shadowing effects resulting an overall shadowing in 30 A GeV Au+Au collisions. In addition the magnitude of the final state nuclear dissociation also increases with decreasing collision energy. Explicit evaluation of the shadowing factots in absence of any final state dissociation ($R_{AA} (\epsilon^{2} = 0)$) indicates that around $15 \%$ reduction of the primordial production cross section arises due to the initial state shadowing effects; rest $75 \%$ can be attributed to the final state nuclear dissociation resulting an overall about $90\%$ reduction of the charmonium yield compared to $p+p$ collisions in central Au+Au collisions, due to nuclear effects alone. Remaining $10 \%$ surviving the cold matter suppression, would then encounter the secondary medium subsequently produced in the collisions. Depending on the plasma characteristics, Debye screening would then dissolve $15 - 20 \%$ of these remaining $J/\psi$ traversing through the medium. One can take note that the differences of $R_{AA}$ with and without plasma screening effects are very little. In the deconfined medium, the contribution to the color screening mass from quarks is dominant and that from gluons is small at low temperatures. At high temperatures, the contributions to the screening mass come from both quarks and gluons and thus color screening effects would be more stronger, higher be the temperature of the plasma. However the plasma anticipated to be produced at FAIR, will have smaller temperatures and higher baryon densities compared to that at SPS or RHIC. Consequently the screening effects will be weaker resulting small suppression due to plasma screening. Hence experimental distinction of the exclusive plasma effects in turn demans very accurate data set, with lowest possible systematic and statistical errors, to be collected at FAIR.

In summary, we have studied the charmonium suppression, in a baryon rich QGP using a simple geometrical screening model. Such high density plasma is believed to exist in the core of the neutron stars and expected to be formed in the low energy nuclear collisions at FAIR. Suppression pattern resulting from color screening is found to be sensitive to the plasma characteristics. Debye mass, as a function of $T$ and $\mu_B$ is used to implement the screening conditions. This implicitly assumes that the nuclear collisions at FAIR create favorable conditions for the formation of a partonic medium in thermal and chemical equilibrium. Chemical equilibration is assumed to occur simultaneously with thermal equilibrium which allows us to estimate the screening mass in terms of $T$ and $\mu_B$. However in practice equilibration might occur in succession with thermalization preceding over the chemical equilibrium. Screening could also be operative in such a thermally equilibrated but chemically equilibrating plasma. Choice of QGP formation time is based on a geometric criterion. Once the two colliding baryon currents have separated again, the system is assumed to be in instantaneous local equilibrium. Instead of screening mass, alternatively one could as well have used threshold dissociation temperature ($T_d$) or energy density ($\epsilon_d$) for demarcation of the screening zone. Though widely used at higher energies, this later approach is particularly disadvantageous at FAIR regime, as the $\mu_B$ dependence of $T_d$ is still not available from the existing lattice calculations. More precise estimations can be made once the screening studies at finite density are available from state-of-the-art lattice simulations. Dissolution of a state due to screening corresponds to dynamical process of the divergence of the thermal deacy widths of the chamonium states in their direct decay channel. However screening results in weakening of the in-medium binding energies of the charmonia and hence they can also get dissociated by inelastic collisions with the surrounding partons co-moving with the evolving charmonium states (see~\cite{Rapp} and references therein). Such partonic dissociation mechanisms could be operative on the bound states as well their precursors and do not require equilibration to set in. They can be operational in the pre-equilibrium stage as well, as along as the medium remains in deconfined state. Collisional dissociation by hard partons has been found to be important at higher energies with energetic gluons playing the dominant role. At FAIR energies, valence quarks rather than gluons will be the prevalent degrees of freedom. Nevertheless quantitative assesments of these effects are necessary. Work is under progress towards the estimation such partonic dissociations and will be communicated in furure. One advantage of charmonium measurements at FAIR, is that the initial suppression effects would possibly not get compensated by subsequent regeneration effects. Because of the larger formation time for any secondary medium, $J/\psi$s during their evolution will mostly encounter the colliding nuclear matter. The cold nuclear effects, which possibly leads to the largest contribution to the overall suppression in nuclear collisions, can be reliably estimated from the high precision data from the $p+A$ collisions. In Au+Au collisions, any additional suppression, can then be used to determine the interplay of the various partonic dissociation mechanisms.


\end{document}